\documentclass[pra,twocolumn,floatfix,a4paper,superscriptaddress]{revtex4}
\usepackage{bm,color,graphicx,amsmath,txfonts}

\newcommand{\FF}{{\cal F}}

\begin{document}

\title{Remote magnon entanglement between two massive ferrimagnetic spheres \\ via cavity optomagnonics}

\author{Wei-Jiang Wu}
\affiliation{Interdisciplinary Center of Quantum Information, Zhejiang Province Key Laboratory of Quantum Technology and Device, and State Key Laboratory of Modern Optical Instrumentation, Department of Physics, Zhejiang University, Hangzhou 310027, China}
\author{Yi-Pu Wang}
\affiliation{Interdisciplinary Center of Quantum Information, Zhejiang Province Key Laboratory of Quantum Technology and Device, and State Key Laboratory of Modern Optical Instrumentation, Department of Physics, Zhejiang University, Hangzhou 310027, China}
\author{Jin-Ze Wu}
\affiliation{Interdisciplinary Center of Quantum Information, Zhejiang Province Key Laboratory of Quantum Technology and Device, and State Key Laboratory of Modern Optical Instrumentation, Department of Physics, Zhejiang University, Hangzhou 310027, China}
\author{Jie Li}\thanks{jieli6677@hotmail.com}
\affiliation{Interdisciplinary Center of Quantum Information, Zhejiang Province Key Laboratory of Quantum Technology and Device, and State Key Laboratory of Modern Optical Instrumentation, Department of Physics, Zhejiang University, Hangzhou 310027, China}
\author{J. Q. You}
\affiliation{Interdisciplinary Center of Quantum Information, Zhejiang Province Key Laboratory of Quantum Technology and Device, and State Key Laboratory of Modern Optical Instrumentation, Department of Physics, Zhejiang University, Hangzhou 310027, China}

\begin{abstract}
Recent studies show that hybrid quantum systems based on magnonics provide a new and promising platform for generating macroscopic quantum states involving a large number of spins. Here we show how to entangle two magnon modes in two massive yttrium-iron-garnet (YIG) spheres using cavity optomagnonics, where magnons couple to high-quality optical whispering gallery modes supported by the YIG sphere. The spheres can be as large as 1 mm in diameter and each sphere contains more than $10^{18}$ spins. The proposal is based on the asymmetry of the Stokes and anti-Stokes sidebands generated by the magnon-induced Brillouin light scattering in cavity optomagnonics. This allows one to utilize the Stokes and anti-Stokes scattering process, respectively, for generating and verifying the entanglement. Our work indicates that cavity optomagnonics could be a promising system for preparing macroscopic quantum states.
\end{abstract}

\date{\today}
\maketitle

\section{introduction}

During the past decade, cavity magnonics~\cite{S1,S2,S3,S4,S5,S6} has been emerged and developed as a new and active platform for the study of strong interactions between light and matter~\cite{Solano}. It consists of microwave photons which reside in a resonant cavity and interact with magnons (i.e., collective spin excitations) in a ferrimagnetic material, e.g., yttrium iron garnet (YIG). The system exhibits its unique features and advantages, which lie in the large frequency tunability and low damping rate of the magnon mode, as well as its excellent ability to coherently interact with other systems, including microwave~\cite{S1,S2,S3,S4,S5,S6} or optical photons~\cite{Usami16,Tang16,Haigh,PRB16,Usami18}, phonons~\cite{CMM,JiePRL18,JiePRL20}, and superconducting qubits~\cite{qubit1,qubit2,qubit3}. These hybrid cavity magnonic systems promise potential applications in quantum information processing and quantum sensing~\cite{NakaRev}. A variety of interesting phenomena have been explored in cavity magnonics, including magnon gradient memory~\cite{Tang15}, exceptional points~\cite{You17,ESurf}, manipulation of distant spin currents~\cite{Hu17}, bi-\cite{yp18} and multi-stability~\cite{yp21}, level attraction~\cite{Xia18,Hu18,Bhoi19,Xiao}, nonreciprocity~\cite{yp19}, anti-PT symmetry~\cite{Yang20,Du20}, among others. In addition, it has been suggested that cavity-magnon polaritons could be used as an ultra-sensitive magnetometer~\cite{Yan,Ruoso,Ali}, and for searching dark matter axions~\cite{DM} and detecting high-frequency gravitational waves using the gravitomagnetic effect~\cite{Matteo}.

In this article, we study an emerging field of cavity optomagnonics~\cite{Usami16,Tang16,Haigh,Usami18}, where a YIG sphere simultaneously supports optical whispering gallery modes (WGMs) and a magnetostatic mode of magnons. The WGM photons are scattered by the GHz magnons in the form of generating sideband photons with the frequency shifted by the magnon frequency, and the high-quality WGM cavity drastically enhances this magnon-induced Brillouin light scattering (BLS). The nature of the spin-orbit coupling of the WGM photons, combined with the geometrical birefringence of the WGM resonator, leads to a pronounced nonreciprocity and asymmetry in the Stokes and anti-Stokes sidebands generated by the magnon-induced BLS~\cite{Usami16,Tang16,Haigh,Usami18}. This is the result of the selection rule~\cite{GEB17,Papa,UsamiNJP,Haigh18} imposed by the angular momentum conservation. Because of this, this kind of BLS requires a change in optical polarization, distinctly different from the light scattering in cavity optomechanics~\cite{OMrmp}. The asymmetry nature of the BLS allows us to select, on demand, the Stokes or anti-Stokes scattering event to occur, corresponding to the process of creating or annihilating magnons. Such a mechanism can be used for the manipulation of magnons, and has been adopted for preparing nonclassical states of magnons. This includes the proposals for cooling the magnons~\cite{GEB18}, preparing magnon Fock states~\cite{SVK19}, a magnon laser~\cite{Xiong}, an optomagnonic Bell test~\cite{Bell}, and an opto-microwave entanglement mediated by magnons~\cite{Zhou}, etc.

Based on this novel system and its unique properties, we provide a scheme to entangle two magnon modes in two massive YIG spheres, which can be separated remotely, manifesting the {\it nonlocal} nature of the macroscopic entanglement.  Specifically, a {\it weak} laser pulse with a certain polarization is sent into an optical interferometer formed by two 50/50 beam splitters (BSs). Each arm of the interferometer contains one cavity optomagnonic device, i.e., a YIG sphere supporting optical WGMs and a magnon mode~\cite{Usami16,Tang16,Haigh,Usami18}, and the magnon mode is cooled to its ground state using a dilution refrigerator. The pulse is detuned to be resonant with the WGM cavities in the two arms and to activate the Stokes scattering event, which yields a single magnon residing in one of the YIG spheres and a lower-frequency photon with a changed polarization in the same arm. Since the BSs are 50/50 and the two devices are assumed identical, the probabilities of the Stokes scattering event in each arm are thus equal. A single-photon detection in the output of the interferometer then projects the two magnon modes onto a path-entangled state, in which the two YIG spheres share a single magnon excitation. 

The scheme can be regarded as the DLCZ protocol~\cite{DLCZ} applied to the system of cavity optomagnonics. The DLCZ protocol for a cavity optomechanical system has been realized using GHz mechanical resonators~\cite{mechEn}. We would like to note that the entanglement between two magnon modes of massive ferrimagnets has been extensively studied in cavity magnonics~\cite{Jie19,GSA19,Yung,GEBB,Jaya,JieJPB,Qian}. Such entangled states involving a large number of spins are genuinely macroscopic quantum states, and are thus useful for the study of the quantum-to-classical transition and the test of unconventional decoherence theories~\cite{Bassi}. In most studies, the magnon entanglement essentially originates from the nonlinearity of the system, which can be achieved from, e.g., the magnetostrictive interaction~\cite{Jie19}, the magnon Kerr effect~\cite{GSA19}, or the coupling to a superconducting qubit~\cite{Qian}. Alternatively, entanglement can be obtained by feeding a squeezed vacuum microwave field into the cavity~\cite{Jaya,JieJPB}. However, by now proposals for entangling two magnon modes by means of cavity optomagnonics are still missing. The magnon entanglement achieved in the present work utilizes the nonlinearity of the magnon-induced BLS, which is a unique property of the cavity optomagnonic system.

\section{Basic interactions in optomagnonics}\label{basic}

We start with the description of two basic interactions in cavity optomagnonics, which are key elements for realizing our protocol. They are the optomagnonic two-mode squeezing and beamsplitter interactions, which are used, respectively, to prepare and verify the magnon entanglement of two YIG spheres. 

The magnon-induced BLS in a cavity optomagnonic system~\cite{Usami16,Tang16,Haigh,Usami18} is intrinsically a three-wave process, which can be described by the Hamiltonian
\begin{equation}
H= H_0 + H_{\rm int},
\end{equation}
where $H_0$ is the free Hamiltonian of two WGMs and a magnon mode
\begin{equation}
H_0/\hbar= \omega_1 a_1^{\dag} a_1 + \omega_2 a_2^{\dag} a_2 + \omega_m m^{\dag} m,
\end{equation}
with $a_j$ and $m$ ($a_j^{\dag}$ and $m^{\dag}$, $j=1,2$) being the annihilation (creation) operators of the WGMs and magnon mode, respectively, and $\omega_i$ ($i=1,2,m$) being their resonance frequencies, which satisfy the relation $\omega_m \ll \omega_j$ and $|\omega_1-\omega_2|=\omega_m$, imposed by the conservation of energy in the BLS. The interaction Hamiltonian $H_{\rm int}$ of the three modes is given by
\begin{equation}
H_{\rm int}/\hbar= G_0 \big(a_1^{\dag} a_2 m^{\dag} + a_1 a_2^{\dag} m \big),
\end{equation}
where $G_0$ is the single-photon coupling rate. This coupling is weak owing to the large frequency difference between the WGM and the magnon mode, but it can be significantly enhanced by intensely driving one of the WGMs. To maximize the BLS scattering probability, we resonantly pump the WGM $a_1$ ($a_2$) to activate the anti-Stokes (Stokes) scattering, which is responsible for the optomagnonic state-swap (two-mode squeezing) interaction. Note that the selection rule~\cite{GEB17,Papa,UsamiNJP,Haigh18} causes different polarizations of the two WGMs. Without loss of generality, we assume $a_2$ ($a_1$) mode to be the transverse-magnetic (TM) (transverse-electric (TE)) mode of a certain WGM orbit, and $\omega_{2({\rm TM})}>\omega_{1({\rm TE})}$ due to the geometrical birefringence of the WGM resonator~\cite{Usami16}.

We now consider that the WGM $a_2$ is resonantly pumped by a strong optical field. In this case, the strongly driven mode $a_2$ can be treated classically as a number $\alpha_2\equiv \langle a_2 \rangle \,{=} \sqrt{N_2}$ ($\alpha_2$ being real for a resonant drive), with $N_2$ the intra-cavity photon number, which is determined by the pump power and the decay rate of the WGM. The linearized interaction Hamiltonian can then be obtained
\begin{equation}
H^{\rm St.}_{\rm int}= \hbar G_2 \big( a_1^{\dag} m^{\dag}  + a_1 m \big),
\end{equation}
where $G_2=G_0 \alpha_2$ is the effective coupling rate. This Hamiltonian is responsible for the two-mode squeezing interaction between the WGM $a_1$ and magnon mode $m$, and can be used to prepare optomagnonic entangled states. This corresponds to the Stokes scattering process, where a TM polarized photon converts into a lower-frequency TE polarized photon by creating a magnon excitation. Under this Hamiltonian, the WGM $a_1$ and magnon mode $m$ are prepared in a two-mode squeezed state (unnormalized)
\begin{equation}\label{eqTMS}
| \psi \rangle_{\rm optomag}= |00 \rangle_{a_1,m} + \!\! \sqrt{P} |11 \rangle_{a_1,m} + P |22 \rangle_{a_1,m} + {\cal O}(P^{3/2}) \,\, ,
\end{equation}
where $P$ is the probability for a single Stokes scattering event to occur, and ${\cal O}(P^{3/2})$ denotes the terms with more excitations whose probabilities are equal to or smaller than $P^3$. The scattering probability $P$ increases with the strength of the driving field. For a sufficiently weak driving field, $P \ll 1$ can be achieved, e.g., in analogous cavity optomechanical experiments, $P \,\,{\simeq}\,\, 0.7\%$~\cite{mechEn} and $P \,\,{\simeq}\,\, 3\%$~\cite{Simon16} were achieved using very weak laser pulses. In this case, the probability of creating two-magnon/photon state $|2 \rangle_{m/a}$ and higher excitation states is negligibly small. Such a low Stokes scattering probability of generating an entangled pair of single excitations, accompanied with very weak laser pulses, is vital for realizing the DLCZ(-like) protocols~\cite{DLCZ,mechEn}. This is exactly what we shall utilize and apply to optomagnonics, in Sec.~\ref{prot}, to generate an entangled pair of a single magnon and a TE polarized photon.

Similarly, when mode $a_1$ is resonantly pumped by a strong field, one obtains the following linearized interaction Hamiltonian
\begin{equation}
H^{\rm A.-St.}_{\rm int}= \hbar G_1 \big( a_2 m^{\dag}  + a_2^{\dag} m \big),
\end{equation}
where $G_1=G_0 \alpha_1$ and $\alpha_1 \,{=} \sqrt{N_1}$, with $N_1$ the intra-cavity photon number of WGM $a_1$. This Hamiltonian leads to the state-swap interaction between the WGM $a_2$ and magnon mode $m$, and can be used to read out the magnon state by measuring the created anti-Stokes field $a_2$. This anti-Stokes scattering corresponds to the process where a TE polarized photon converts into a higher-frequency TM polarized photon by annihilating a magnon. 
As will be shown in Sec.~\ref{veri}, we shall use this anti-Stokes process to verify the magnon entanglement.

\begin{figure}[t]
\hskip0.0cm\includegraphics[width=\linewidth]{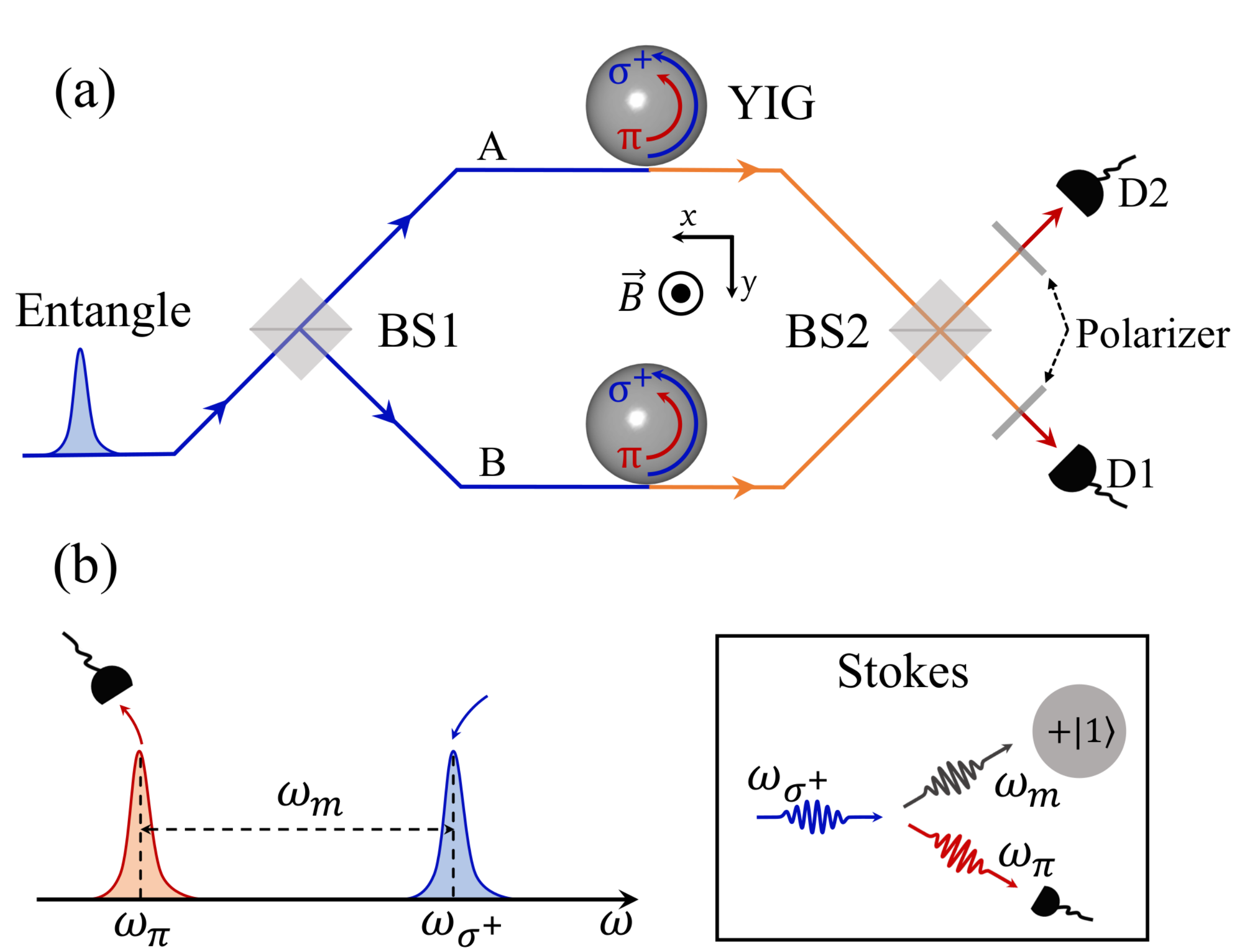} 
\caption{(a) Sketch of the system used for generating the magnon entanglement between two YIG spheres. It consists of an interferometer formed by two 50/50 BSs and in each arm of the interferometer there is a cavity optomagnonic device, in which the magnon-induced Brillouin light scattering occurs. A weak laser pulse with a certain polarization is sent into the interferometer and a subsequent single-photon detection in the output of the interferometer projects the two magnon modes onto an entangled state. (b) Mode frequencies of the Stokes light scattering by magnons. A $\sigma^+$-polarized photon of frequency $\omega_{\sigma^+}$ is converted into a $\pi$-polarized Stokes photon of frequency $\omega_{\pi}$ by creating a magnon of frequency $\omega_m$. The input pulse couples to the $\sigma^+$-polarized WGM and the generated $\pi$-polarized photon goes into the detector, and meanwhile, the magnon mode gains a single excitation. }
\label{fig1}
\end{figure}

\section{The protocol}\label{prot}

We now proceed to describe our protocol. The schematic diagram of the protocol is depicted in Fig.~\ref{fig1}. Two cavity optomagnonic devices are placed in two arms of an optical interferometer formed by two 50/50 BSs. In each device, a YIG sphere supports a magnon mode and high-$Q$ optical WGMs. In Refs.~\cite{Usami16,Haigh,Usami18}, a roughly 1-mm-diameter sphere was used. The YIG sphere is placed in a bias magnetic field along the $z$ direction, while the WGMs propagate along the perimeter of the sphere in the $x$-$y$ plane~\cite{Usami16,Tang16,Haigh,Usami18}. The frequency of the magnon mode can be adjusted by varying the strength of the bias magnetic field. The two devices and two optical paths are assumed identical to erase the `which-device/path' information of the scattered photons at the output of the interferometer (though some mismatches can be compensated via optical operations~\cite{mechEn}), which is required by the DLCZ protocol. As a particular advantage of the system, the two magnon-mode frequencies can be tuned to be equal by altering the external bias fields. This is technically more difficult for optomechanical systems as one has to fabricate a large number of samples and find a pair of nearly identical mechanical resonators~\cite{mechEn}. At the end of the interferometer, two single-photon detectors are placed in the outputs of the 50/50 BS, and in each output a polarizer is placed in front of the detector, to select the photon with a certain polarization to be detected.

We now describe the scheme first by neglecting any optical losses, including the propagation loss and the detection loss due to the nonunity detection efficiency, as well as the loss associated with the magnon modes. Also, we further assume that the two magnon modes are prepared in their quantum ground state. We then analyse the effects of various experimental imperfections, and finally provide a strategy for verifying the entanglement.

A laser pulse is sent into one input port of the first BS of the interferometer. The energy of the pulse is so weak that the mean photon number is much smaller than 1~\cite{DLCZ,mechEn}. It is thus in a weak coherent state $|\alpha \rangle$ ($|\alpha|\ll 1$), with the probability of all the $n$-photon components $|n\rangle$ ($n>1$) being negligible, i.e., $|\alpha \rangle \simeq |0 \rangle +\!\! \sqrt{p} |1 \rangle$, where $p=|\alpha|^2 \ll 1$ is the probability of the pulse being in the single-photon state. Since the BS is 50/50, the single photon goes, with equal probability, into one of the interferometer arms (the two arms are termed as path A and B, see Fig.~\ref{fig1}(a)), thus having the state after the BS
\begin{equation}\label{eq1}
| \phi \rangle_{\rm opt} \simeq |00 \rangle_{\rm AB} + \! \frac{\sqrt{p}}{\sqrt{2}} \big( |01 \rangle_{\rm AB} +  |10 \rangle_{\rm AB} \big).
\end{equation}
The pulse is sent through a fiber polarization controller and coupled to the WGM resonator via a tapered silica optical nanofiber~\cite{Usami16}, or a prism coupler~\cite{Haigh,Usami18}. The polarization and the frequency of the pulse is tuned to couple to a certain TM WGM, e.g., the $\sigma^+$-polarized TM mode. Therefore, the state of the system, soon after the TM mode is excited, is
\begin{equation}\label{eq2}
| \phi' \rangle_{\rm opt} \simeq |00 \rangle_{\sigma^+_A \sigma^+_B} + \! \frac{\sqrt{p}}{\sqrt{2}} \big( |01 \rangle_{\sigma^+_A \sigma^+_B} +  |10 \rangle_{\sigma^+_A \sigma^+_B} \big),
\end{equation}
where the subscript $\sigma^+_j$ ($j\,{=}\,$A, B) denotes the $\sigma^+$-polarized TM mode of the WGM resonator in path $j$. The $\sigma^+$-polarized photon is then scattered by creating a magnon, and generates a $\pi$-polarized Stokes photon at frequency $\omega_{\pi} = \omega_{\sigma^+} - \omega_m$ into the TE WGM, see Fig.~\ref{fig1}(b). This is just what we introduced, in Sec.~\ref{basic}, the low Stokes scattering probability of generating an entangled pair of a single magnon and a TE polarized photon. This magnon-induced Stokes BLS leads to the following state 
\begin{equation}\label{eq3}
\begin{split}
| \phi \rangle_{\rm optomag}  \simeq  & \,\, |0000 \rangle_{m_A m_B \pi_A \pi_B}   \\
& + \! \frac{\sqrt{p}}{\sqrt{2}} \big( |0101 \rangle_{m_A m_B \pi_A \pi_B} +  |1010 \rangle_{m_A m_B \pi_A \pi_B} \big),
\end{split}
\end{equation}
where $ |0101 \rangle_{m_A m_B \pi_A \pi_B}$ denotes the coexistence of a magnon residing in the YIG sphere and a $\pi$-polarized Stokes photon in path B, and similarly for $|1010 \rangle_{m_A m_B \pi_A \pi_B}$. Note that this BLS occurs only between the TM and TE modes with the same WGM index, owing to the angular momentum conservation of photons. Because of the geometrical birefringence, which imposes a restriction on the frequencies of the TM and TE modes (of the same mode index), i.e., $\omega_{\rm TM} > \omega_{\rm TE}$, the Stokes scattering is preferred in the BLS, while the anti-Stokes scattering is prohibited, in which a $\sigma^-$-polarized photon is converted into an anti-Stokes photon at frequency $\omega_{\pi} = \omega_{\sigma^-} + \omega_m$ by annihilating a magnon~\cite{Usami16}. As discussed later, in the part of entanglement verification, we shall send a pulse that is coupled to a TE WGM to activate the anti-Stokes scattering. Such an asymmetry nature of the BLS by the magnons is the cornerstone of realizing our scheme, which offers the possibility of separately implementing the entangling operation and the readout operation of the scheme.

The generated $\pi$-polarized Stokes photon, with equal probability in path A or B, then couples to the nanofiber (or the prism coupler) and enters the second 50/50 BS. The polarizers in the outputs of the BS select the TE Stokes photon over the TM photons that failed to effectively activate the Stokes scattering (in practice, the experiment will be repeated many times due to the low scattering probability in the current optomagnonic weak coupling regime~\cite{Usami16,Tang16,Haigh,Usami18}). A single-photon detection in the output of the BS, which realizes the measurement $M_{\pm}= \big( |01 \rangle_{\pi_A \pi_B} \pm  |10 \rangle_{\pi_A \pi_B} \big)^{\dag}$, then projects the two magnon modes onto the state
\begin{equation}\label{eq4}
| \phi \rangle_{\rm mag} = \! \frac{1}{\sqrt{2}} \big( |01 \rangle_{m_A m_B} \pm  |10 \rangle_{m_A m_B} \big).
\end{equation}
This is a path-entangled state of the two magnon modes in paths A and B, and `$\pm$' correspond to the detection of the TE photon in different outputs of the BS.


In deriving the entangled state~\eqref{eq4}, we have neglected the optical losses (e.g., the propagation loss and the detection loss), and the magnon loss. Also, we have assumed that the magnon modes are initialized to their quantum ground state $|00 \rangle_{m_A m_B}$ by eliminating the residual thermal excitations. We now analyse these effects one after another. 

For both the optical propagation loss and the detection loss due to the nonunity detection efficiency, their effect only reduces the probability of obtaining the desired state~\eqref{eq4} but does not destroy the state~\cite{JieTele}, which implies longer measurement time. This is only true for a sufficiently weak pulse (being in a weak coherent state $|\alpha \rangle$ with $|\alpha|\,{\ll}\, 1$) that produces {\it at most} a single Stokes photon in the output of the interferometer. Optical losses thus disable the single-photon detection. This experiment will be disregarded such that there is no actual impact. However, the two-photon component in $|\alpha \rangle$ can indeed result in unwanted additional states, such as $|02 \rangle_{m_A m_B}$, $|20 \rangle_{m_A m_B}$, $|11 \rangle_{m_A m_B}$, in the final magnon state~\eqref{eq4}. This is also true for the case without suffering any optical loss as the detectors are assumed to be photon-number non-resolving in our scheme. Nevertheless, the probability of the two-photon state is much smaller than that of the single-photon state for a weak coherent state with $|\alpha| \ll 1$. As long as this is satisfied, those additional states are negligible.

For the dissipation of the magnon modes, since the timescale at which our scheme is realized (laser pulses with duration of tens of nanoseconds were used in Refs.~\cite{mechEn,Simon16}) is much shorter than the magnon lifetime (typically of a microsecond~\cite{Usami16,Tang16,Haigh,Usami18}), during a complete run of the experiment the magnon modes can be assumed to have negligible dissipation. However, the magnon modes cannot be perfectly initialized to their ground state $|00 \rangle_{m_A m_B}$ at typical cryogenic temperatures. We now study the impact of the residual magnon thermal excitations. Since the frequencies of the two magnon modes are tuned to be equal, the two magnon modes are in the same thermal state under the same temperature
\begin{equation}\label{eq5}
\rho_{\rm mag}^{\rm th} = (1{-}S) \sum_{n=0}^{\infty} S^n \, | n \rangle \langle n | , 
\end{equation}
where $S\,{=}\,\bar n/(\bar n +1)$, with $\bar n=\big[ {\rm exp} (\hbar \omega_m/k_B T) {-}1 \big]^{-1}$ being the equilibrium mean thermal magnon number at the temperature $T$. In general, quantum states of macroscopic objects require very low environmental temperatures. For the magnon mode with frequency of about 7 GHz~\cite{Usami16,Tang16,Haigh,Usami18}, the thermal occupation $\bar n \simeq 0.036$ at $T=100$ mK. For $\bar n = 0.036$, $S \simeq 0.035$ and $S^2 \,\,{\simeq}\,\, 0.001$, therefore, high-excitation terms $|n \rangle$ with $n\,\,{>}\,\,1$ can be safely neglected, and we can then approximate it as $\rho_{\rm mag}^{\rm th} \simeq (1{-}S)  \big( |0 \rangle \langle 0| + S |1\rangle \langle 1| \big)$. The two magnon modes are thus in a mixed state of a probabilistic mixture of four pure states $| ij \rangle_{m_A m_B}$ ($i\, (j)\,\,{=}\,\,0,1$). The ratio of the probabilities is 1 : $S$ : $S$ : $S^2$ for the two magnon modes being initially in $| 00 \rangle_{m_A m_B}$, $| 01 \rangle_{m_A m_B}$, $| 10 \rangle_{m_A m_B}$, and $| 11 \rangle_{m_A m_B}$, respectively. The ground state $| 00 \rangle_{m_A m_B}$ is what we have assumed for obtaining the desired state $| \phi \rangle_{\rm mag}$ in \eqref{eq4}. Combining the other initial states, we obtain the final magnon state (unnormalized), conditioned on the single-photon detection, which is
\begin{equation}\label{eq6}
\rho_{\rm mag}^{\rm final} \simeq  | \phi_{00} \rangle \langle \phi_{00} | + S \big( | \phi_{01} \rangle \langle \phi_{01} | + | \phi_{10} \rangle \langle \phi_{10} | \big) + S^2 | \phi_{11} \rangle \langle \phi_{11} |, 
\end{equation}
where $ | \phi_{00} \rangle \equiv | \phi \rangle_{\rm mag}$, and 
\begin{equation}\label{eq7}
\begin{split}
| \phi_{01} \rangle &= \! \frac{1}{\sqrt{2}} \big( |02 \rangle_{m_A m_B} \pm  |11 \rangle_{m_A m_B} \big), \\
| \phi_{10} \rangle &= \! \frac{1}{\sqrt{2}} \big( |11 \rangle_{m_A m_B} \pm  |20 \rangle_{m_A m_B} \big), \\
| \phi_{11} \rangle &= \! \frac{1}{\sqrt{2}} \big( |12 \rangle_{m_A m_B} \pm  |21 \rangle_{m_A m_B} \big), \\
\end{split}
\end{equation}
corresponding to the magnon modes being initially in $| 01 \rangle_{m_A m_B}$, $| 10 \rangle_{m_A m_B}$, and $| 11 \rangle_{m_A m_B}$, respectively. The states in \eqref{eq7} reduce the fidelity of the desired state~\eqref{eq4}, as $\FF = \langle \phi_{00}| \, \rho_{\rm mag}^{\rm final} \, |\phi_{00} \rangle \simeq 1/(1{+}2S{+}S^2)$. Nevertheless, these additional states can be well suppressed if $S \,\,{\ll} \,\,1$. This is well fulfilled at the temperature of 100 mK (50 mK), since $S \simeq 0.035 \,\, (0.001) \ll 1$. Under this temperature, the fidelity of the state~\eqref{eq4} is $\FF \simeq 0.93$ (0.998). Therefore, the impact of the residual thermal excitations under a temperature below 100 mK will be negligibly small.

\section{Verification of the entanglement}\label{veri}

\begin{figure}[t]
\hskip0.0cm\includegraphics[width=\linewidth]{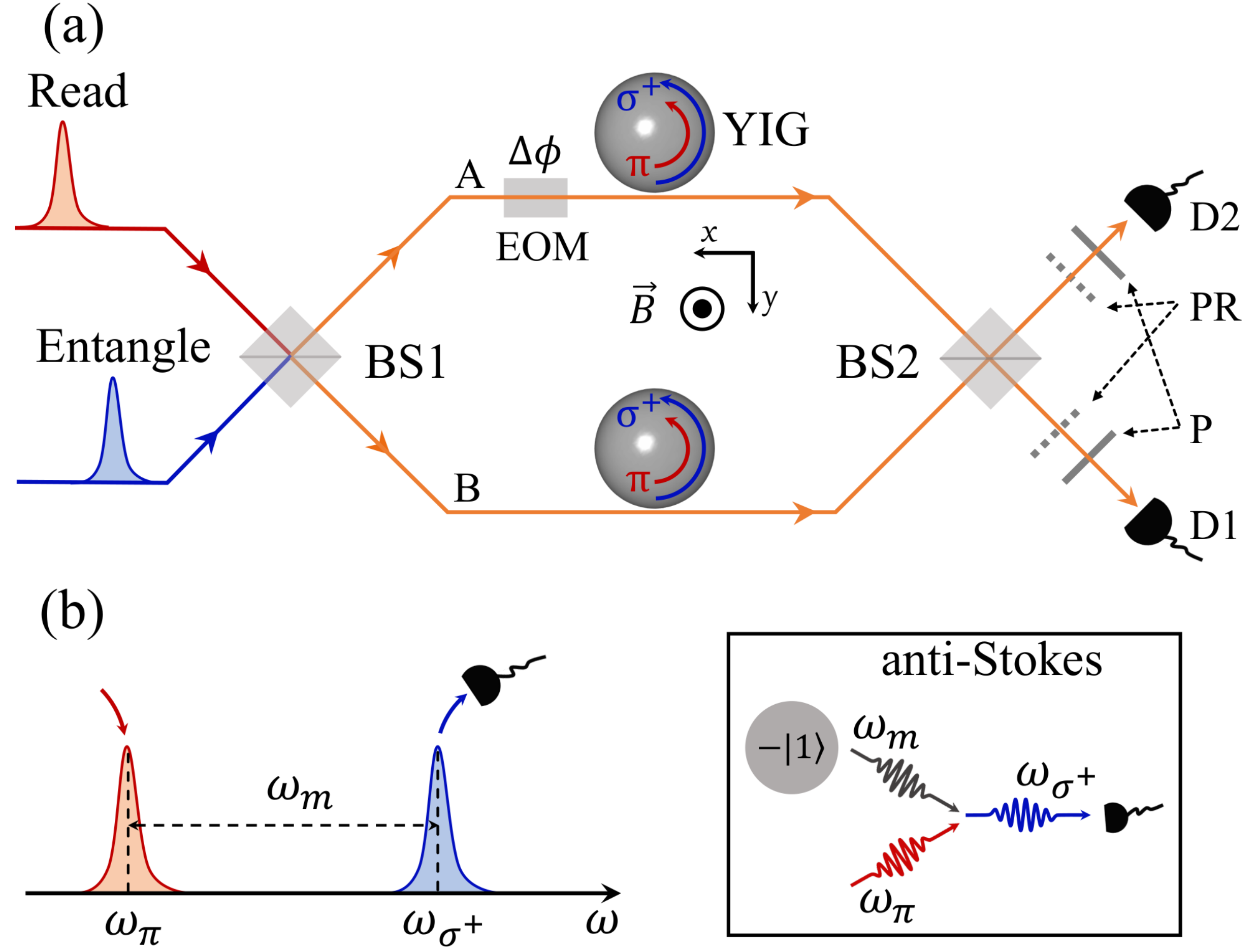} 
\caption{(a) Sketch of the system for verifying the magnon entanglement. A weak read pulse is sent into the interferometer soon after the entangling pulse and couples to a $\pi$-polarized WGM to activate the anti-Stokes scattering. The created $\sigma^+$-polarized anti-Stokes photon, containing the magnon state information, enters one of the detectors after passing through a polarization rotator (PR) and a polarizer (P). An electro-optic modulator (EOM) is added in one arm to realize the phase offset $\Delta\phi$. (b) Mode frequencies of the anti-Stokes light scattering by magnons. A $\pi$-polarized photon is converted into a $\sigma^+$-polarized anti-Stokes photon by annihilating a magnon that was produced in the preceding entangling stage.  }
\label{fig2}
\end{figure}

Lastly, we show how to verify the generated magnon entanglement. The magnon state can be read out by using the anti-Stokes process of the BLS. Specifically,  as depicted in Fig.~\ref{fig2}, a weak read pulse is sent into the interferometer with its polarization and frequency tuned to couple to a TE WGM to activate the anti-Stokes scattering, where a $\pi$-polarized photon is converted into a $\sigma^+$-polarized anti-Stokes photon by annihilating a magnon, satisfying the frequency relation $ \omega_{\sigma^+} = \omega_{\pi} + \omega_m$~\cite{Usami16}. This can be regarded as the inverse process of the Stokes scattering used for preparing the entanglement. The generated $\sigma^+$-polarized anti-Stokes photon then couples to the nanofiber/prism coupler and enters the second BS. In this circumstance, the polarizers in the outputs of the BS select the TM anti-Stokes photon over the TE photons that disable the anti-Stokes scattering. In practice, the read pulse is sent after the entangling pulse with a time delay {\it much shorter} than the magnon lifetime. Within the delay, the polarizers must be quickly switched in order to select the TM photon for verifying the entanglement.  Alternatively, one could keep the polarizer fixed and put a polarization rotator before the polarizer, which quickly rotates the polarization of the photon (TM $\leftrightarrow$ TE). This can be realized via a high-speed waveguide electro-optic polarization modulator~\cite{PR1}. 

In view of the similarity of optomagnonics and optomechanics~\cite{note}, we adopt the following witness for the magnon entanglement~\cite{Girvin,mechEn}
\begin{equation}\label{eq8}
R_m(\Delta\phi, j) = 4 \frac{ g_{A_1,S_j}^{(2)}(\Delta\phi) + g_{A_2,S_j}^{(2)}(\Delta\phi) -1}{ \Big( g_{A_1,S_j}^{(2)}(\Delta\phi) - g_{A_2,S_j}^{(2)}(\Delta\phi)  \Big)^2 },
\end{equation}
where $g_{A_i,S_j}^{(2)}=\langle A_i^{\dag} S_j^{\dag} A_i S_j \rangle/\big( \langle A_i^{\dag} A_i \rangle \langle S_j^{\dag} S_j \rangle \big)$ is the second-order coherence between the TE Stokes photons (with $S_j$ and $S_j^{\dag}$ the annihilation and creation operators for the Stokes photons going to detector $j$, $j\,{=}\,1,2$) and the TM anti-Stokes photons (with $A_i$ and $A_i^{\dag}$ the annihilation and creation operators for the anti-Stokes photons going to detector $i$, $i\,{=}\,1,2$), and $\Delta\phi$ is the phase offset added to the read pulse in one of the interferometer arms via an electro-optic modulator (EOM), see Fig.~\ref{fig2}. The witness gives $R_m(\Delta\phi, j) \ge 1$ for all separable states of the two magnon modes for any $\Delta\phi$ and $j$. Therefore, if there is any $\Delta\phi$ and $j$ with which $R_m(\Delta\phi, j) < 1$, the two magnon modes are then entangled.

\section{conclusion}

We present a scheme for entangling two magnon modes of two massive ferrimagnetic spheres in an optical interferometer configuration, containing two optomagnonic devices, by using short optical pulses in a cryogenic environment. The scheme is based on the asymmetry of the Stokes and anti-Stokes sidebands in the magnon-induced Brillouin light scattering. The entanglement is generated on the condition of the detection of single photons with a certain polarization. We analyse the effects of various experimental imperfections and provide a strategy for verifying the entanglement based on the second-order coherence between the Stokes and anti-Stokes photons with different polarizations. The magnon entangled state of two ferrimagnetic spheres, which contain more than $10^{18}$ spins and can be distantly separated, is truly a macroscopic quantum state and manifests its nonlocal nature. Our work, along with the earlier studies~\cite{GEB18,SVK19,Bell}, show that cavity optomagnonics could become a promising platform for the study of macroscopic quantum phenomena.

\section*{Acknowledgments}

This work has been supported by Zhejiang Province Program for Science and Technology (Grant No. 2020C01019), the National Natural Science Foundation of China (Grants Nos. U1801661, 11934010, 11774022 and 12004334), the Fundamental Research Funds for the Central Universities (No. 2021FZZX001-02), and China Postdoctoral Science Foundation.


\begin{thebibliography}{99}
\bibitem{note}
The optomagnonic interaction is formally equivalent to the optomechanical interaction (thus radiation pressure-like)~\cite{Flatte,SVK16}, under the assumption of the low-lying excitations, which allows treating the collective motion of the spins as a harmonic oscillator described by boson operators.
\end{thebibliography}
\end{document}